# Modulation of Schottky barrier height in graphene/MoS$_2$/metal vertical heterostructure with large current ON-OFF ratio

Yohta Sata[1], Rai Moriya[1,*], Takehiro Yamaguchi[1], Yoshihisa Inoue[1], Sei Morikawa[1], Naoto Yabuki[1], Satoru Masubuchi[1,2], and Tomoki Machida[1,2,*]

[1] *Institute of Industrial Science, University of Tokyo, Meguro, Tokyo 153-8505, Japan*
[2] *Institute for Nano Quantum Information Electronics, University of Tokyo, Meguro, Tokyo 153-8505, Japan*

Detail transport properties of graphene/MoS$_2$/metal vertical heterostructure have been investigated. The van der Waals interface between the graphene and MoS$_2$ exhibits Schottky barrier. The application of gate voltage to the graphene layer enables us to modulate the Schottky barrier height; thus gives rise to the control of the current flow across the interface. By analyzing the temperature dependence of the conductance, the modulation of Schottky barrier height $\Delta\varphi$ has been directly determined. We observed significant MoS$_2$ layer number dependence of $\Delta\varphi$. Moreover, we demonstrate that the device which shows larger $\Delta\varphi$ exhibits larger current modulation; this is consistent with the fact that the transport of these devices is dominated by graphene/MoS$_2$ Schottky barrier.

E-mail: moriyar@iis.u-tokyo.ac.jp; tmachida@iis.u-tokyo.ac.jp

## 1. Introduction

After the discovery of graphene: just one atomic layer of carbon with hexagonal crystalline lattice[1-4], researches are boosted in the field of two-dimensional crystals and their van der Waals heterostructures[5-9]. Among them, the van der Waals heterostructures consisting of graphene and semiconducting transition metal dichalcogenides (TMD) such as $MoS_2$ and $WS_2$ are subject of considerable interest for both fundamental and application point of view. It has been shown that the interface between graphene and semiconducting TMD exhibits Schottky barrier; similarly to the conventional metal/semiconductor interface. However, unconventionally, the Schottky barrier height can be strongly modulated by external gate voltage owing to graphene's small density of states[10-16]. Based on this mechanism, various types of vertical heterostrucure transistors have been demonstrated[17-25]. Recently, we demonstrated large ON-OFF ratio of $10^5$ and large ON current density of $10^4$ $A/cm^2$ in exfoliated-graphene/$MoS_2$/Ti vertical field-effect transistor (FET)[26]. Such high performance of vertical heterostructure opens up new possibility of using van der Waals heterostructure for electronics and optoelectronics applications. However, up to now, it has not been fully understood what is the origin of large ON-OFF ratio in exfoliated-graphene/$MoS_2$/Ti.

To address this point, here we show the detail analysis on the temperature dependence of the vertical transport in exfoliated-graphene/$MoS_2$/Ti vertical heterostructures. We observed strong temperature dependence of the device conductance, which is due to the thermionic emission current across the Schottky barrier. By analyzing the Arrhenius plots of the conductance, the modulation of Schottky barrier height $\Delta\varphi$ has been directly determined. We observed significant $MoS_2$ layer number dependence of $\Delta\varphi$. Further, we demonstrate that the device which shows larger $\Delta\varphi$ exhibits larger current ON-OFF ratio. These results suggest that the thermionic emission current across the graphene/MoS$_2$ interface is crucial role for optimal performance of vertical FETs, and reveal that the large $\Delta\varphi$ obtained with exfoliated graphene gives rise to large ON-OFF ratio.

## 2. Experiment

The schematic illustration of the graphene/MoS$_2$/metal heterostructure is shown in Fig. 1(a). By using mechanical exfoliation of Kish graphite, a single layer graphene is fabricated on 300 nm SiO$_2$/$n$-Si(100) substrate; here highly doped Si substrate can be used as a back gate electrode with the SiO$_2$ as a gate dielectric. Next a MoS$_2$ with layer number $N = 14 - 62$ is mechanically cleaved from bulk MoS$_2$ crystal (2D semiconductors Inc.) and deposited on the graphene using dry transfer method[27,28]. This method enables us to create van der Waals interface between freshly cleaved surface of graphene and MoS$_2$. The junction areas for series of devices are ranged $1 - 3$ μm$^2$. By using standard electron-beam (EB) lithography and EB evaporation, the Au 30 nm/ Ti 50 nm electrodes are fabricated on both graphene and MoS$_2$. The current-voltage characteristics of vertical heterostrucrure are measured by applying bias voltage $V_B$ in between graphene and MoS$_2$. The back gate voltage in the rage of $V_G = -50$ to $+50$ V is applied to the highly doped Si substrate to change the graphene's carrier concentration.

## 3. Results and discussion

The band structure at the graphene/MoS$_2$ interface can be schematically illustrated as shown in Fig. 1(b) for both hole-doped ($V_G = -50$ V) and electron-doped ($V_G = +50$ V) cases. At $V_G = -50$ V, the Fermi level of the graphene is the lowest; thus the Schottky barrier at graphene/MoS$_2$ is the highest. In this case, the transport is dominated by the Schottky barrier at graphene/MoS$_2$ interface and exhibits high resistance state; thereby OFF state. Here we assume the Schottky barrier at MoS$_2$/Ti interface is lower than that of graphene/MoS$_2$ interface. From the in-plane transport property of a MoS$_2$, this barrier height is determined as 0.13 eV. On the other hand, at $V_G$ = +50 V, the Fermi level of the graphene is the highest; thus the Schottky barrier height is the lowest. The graphene/MoS$_2$ interface is nearly ohmic and therefore the vertical transport of the device is low resistance state; thus ON state. By switching between the ON and OFF states with $V_G$, the device can be operated as vertical FET.

First, the current-voltage (*I-V*) characteristics are measured at different $V_G$ in graphene/MoS$_2$/Ti vertical heterostructures with the number of MoS$_2$ layers of *N* = 44 and 14. The results are presented in Figs. 2(a) and 2(b). The *I-V* curve is normalized by its junction area of 1.0 µm$^2$ for *N* = 44 and 1.2 µm$^2$ for *N* = 14, respectively. In both cases, the *I-V* curves are significantly modulated with $V_G$. The conductance is the smallest at $V_G$ = -50 V and monotonically increases as $V_G$ is swept toward $V_G$ = +50 V. We revealed that by sweeping back gate voltage, the *I-V* curves can be continuously modulated between these two states. The *I-V* curves for $V_G$ = -50 V show current rectification. On the other hand, the *I-V* curves for the $V_G$ = +50 V show the symmetric structures with respect to $V_B$. These behaviors are consistent with the schematic band diagram presented in Fig. 1(b). A current density vs. $V_G$ is plotted at constant $V_B$ = +0.5 V for both devices as shown in Fig. 2(c). The ON state current density is nearly the same for both devices and this value is determined by the series resistance of the devices rather than that of graphene/MoS$_2$ Schottky barrier[17,26]. On the other hand, the OFF state current density is significantly different between two devices. We achieved much lower OFF current density for *N* = 44 than $N = 14$. We defined ON-OFF ratio as $I(V_G = +50$ V$)/I(V_G = -50$ V$)$ at fixed $V_B = +0.5$ V. The large ON-OFF ratio of $5\times10^4$ is obtained for $N = 44$. By using exfoliated graphene and fabricating high quality interface between graphene and MoS$_2$ by using dry transfer technique, we obtained large ON-OFF ratio in our vertical FETs. The ON-OFF ratio is smaller in thinner MoS$_2$. Such thickness dependences seem to be a general behavior for this type of vertical FETs and have been also observed in our previous reports[17,22,26].

Next, the temperature dependences of the vertical FETs have been measured. The relation between current density and $V_G$ is measured at different temperature from 300 K to 160 K at the fixed $V_B = +0.1$ V. Results are plotted for both $N = 44$ and 14 in Figs. 3(a) and 3(b), respectively. The current monotonically decreases with decreasing temperature. Such significant temperature dependence suggests that the conduction through the device is due to the thermally activated process. The device with $N = 44$ exhibited much stronger temperature dependence than the device with $N = 14$. To analyze this in detail, the $\ln(I/T^2)$ is plotted against $1000/T$ for $V_G = +50$ and $-50$ V, respectively in Figs. 3(c) and 3(d). The obtained curves exhibited straight line. From thermionic emission theory, the slopes of these curves are given by $-\varphi/k_B$, where $\varphi$ denotes the Schottky barrier height and $k_B$ the Boltzmann's constant[29]. At $V_G = +50$ V (ON state), as shown in Fig. 3(c), the slopes of the both curves are almost zero; suggesting that the absence of potential barrier in both devices. At $V_G = -50$ V (OFF state), as shown in Fig. 3(d), the slope of the curve is larger for $N = 44$ than $N = 14$; thus suggesting larger $\varphi$ for $N = 44$. Series of fitting procedures are performed at various $V_G$ and extracted relationship between $\varphi$ and $V_G$ is presented in Fig. 3(e). The $\varphi$ is strongly modulated by the $V_G$. As $V_G$ is swept from -50 to

+50 V, the $\varphi$ decreases; consistent with the graphene/MoS$_2$ Schottky model illustrated in Fig. 1(b). The relationship between $\varphi$ and $V_G$ shows large changes around the Dirac point of the graphene layer. The Dirac point of the graphene layer is $V_G = 0$ V for $N = 44$ and $V_G = -10$ V for $N = 14$, respectively. We think this is related to the change of graphene's Fermi level $E_F$ on $V_G$. Analytically, $E_F$ is expressed as[30]

$$E_F = \text{sign}(V_G - V_{DP})\hbar v_F \sqrt{\pi\alpha(V_G - V_{DP})}, \qquad (1)$$

where $\hbar$ denotes the Planck constant, $v_F = 1 \times 10^6$ m/s the graphene's Fermi velocity, and $\alpha = 7.1 \times 10^{10}$ cm$^{-2}$V$^{-1}$ the capacitance of 300-nm-thick SiO$_2$ gate dielectrics in electron charge. Thus, the change of $E_F$ with respect to $V_G$ is the largest around $V_{DP}$. The potential barrier height is close to zero around $V_G = +50$ V and this suggests nearly ohmic-like behavior as we illustrated in Fig. 1(b). We defined the change of barrier height $\Delta\varphi$ as $\Delta\varphi = \varphi(V_G = -50$ V$) - \varphi(V_G = +50$ V$)$ and obtained $\Delta\varphi$ of around 70 and 240 meV for $N = 14$ and 44, respectively. We obtained larger $\Delta\varphi$ for thicker MoS$_2$.

The temperature dependence of the conductance are measured on the graphene/MoS$_2$/Ti device with different MoS$_2$ thickness ranged $N = 14 - 62$. From series of measurements, the $\Delta\varphi$ and ON-OFF ratios are obtained for different $N$ and results are summarized in Figs. 4(a) and 4(b). The both $\Delta\varphi$ and ON-OFF ratio increase with $N$ and tend to saturate above $N \sim 20$. The $N$ dependence of the Schottky barrier height $\Delta\varphi$ determined from Arrhenius plot and ON-OFF ratio of the graphene/MoS$_2$/Ti vertical FET exhibits good coincidence each other. These results provide a piece of evidence that the modulation of the Schottky barrier height at graphene/MoS$_2$ interface plays dominant role for the current ON-OFF operation of vertical FET. Note that the reduction of ON-OFF ratio in smaller $N$ is also reported by several previous reports and is attributed to the metal-induced Schottky barrier height lowering[17,26]. The metal-induced Schottky barrier height lowering can be attributed to two different origins; these are image force lowering and quantum capacitance of graphene. Firstly, due to the image force lowering effect of Schottky barrier, the effective barrier height becomes lower than ideal value[29]. The reduction of barrier height is proportional to $\sqrt{E}$, where $E$ denotes the electric field inside of MoS$_2$. Since the thickness range of MoS$_2$ we studied is smaller than its depletion width, we could reasonably assume $E \propto 1/d$, where $d$ denotes the thickness of MoS$_2$; therefore this contribution is proportional to $1/\sqrt{d}$. In consequence, in thinner MoS$_2$ devices Schottky barrier height decreases significantly and as a result $\Delta\varphi$ is reduced. Secondly, due to the small density of states of graphene layer, the quantum capacitance becomes non-negligible contribution in the graphene-based vertical heterostructure[31,32]. This contribution reduces the efficiency of the gate electric field effect and thus reduces the $\Delta\varphi$. The reduction of $\Delta\varphi$ is roughly proportional to $\sqrt{C_{MoS_2}} = \sqrt{\varepsilon_{MoS_2}/d}$, where $C_{MoS_2}$ denotes the capacitance per unit area of MoS$_2$ and $\varepsilon_{MoS_2}$ the dielectric constant of MoS$_2$. Therefore, this contribution is also proportional to $1/\sqrt{d}$. Both image force and quantum capacitance contribution become increasingly effective in thinner MoS$_2$ and reduce $\Delta\varphi$; these reasons explain overall features presented in Fig. 4(a).

For the thermionic emission at the graphene/MoS$_2$ interface, the ON-OFF ratio under the sweep of $V_G$ at given temperature $T$ can be simply given by $\exp(\Delta\varphi/k_B T)$. Thus, we plotted $\Delta\varphi$ vs ON-OFF ratio as shown in Fig. 4(c). Clearly, the ON-OFF ratio increases as Δ$\varphi$ increases. The dashed line in the figure represents ON-OFF ratio = exp(Δ$\varphi$/0.02585) at $T$ = 300 K. The experimental data shows good agreement with this relationship. Figure 4(c) suggests another interesting aspect of graphene-based vertical FETs. The larger the Δ$\varphi$ becomes, the larger ON-OFF ratio can be obtained. According to Eq. (1) the maximum modulation of the graphene's Fermi level between $V_G$ = -50 and +50 V with 300 nm SiO$_2$ gate dielectrics is 440 meV[30]. Therefore by carefully optimizing structure or improving quality of the graphene/MoS$_2$ interface, we expect Δ$\varphi$ can be increased as large as 440 meV; this give rises to ON-OFF ratio on the order of $10^7$.

## 4. Summary

Detail transport properties of graphene/MoS$_2$/metal vertical heterostructure have been investigated. By analyzing the temperature dependence of the conductance, the modulation of Schottky barrier height Δ$\varphi$ has been directly determined. We observed significant MoS$_2$ layer number dependence of Δ$\varphi$. Moreover, we demonstrated that the device which showed larger Δ$\varphi$ exhibited larger current modulation; this is consistent with the fact that the transport of these devices is dominated by graphene/MoS$_2$ Schottky barrier.

**Acknowledgements**

This work was partly supported by Grants-in-Aid for Scientific Research from the Japan Society for the Promotion of Science (JSPS); the Science of Atomic Layers, a Grant-in-Aid for Scientific Research on Innovative Areas from the Ministry of Education, Culture, Sports, Science and Technology (MEXT), Japan; and the Project for Developing

Innovation Systems of MEXT. S. Morikawa acknowledges the JSPS Research Fellowship for Young Scientists.

**Figure captions**

Figure 1. (Color online)
(a) Schematic illustration of the graphene (Gr)/MoS$_2$/Ti vertical FET. (b) Band alignment of graphene/MoS$_2$/Ti vertical FET at OFF-state ($V_G$ = -50 V) and ON-state ($V_G$ = +50 V).

Figure 2. (Color online)
(a, b) *I–V* characteristics of the graphene/MoS$_2$/Ti vertical FET under the application of various $V_G$ measured at 300 K with different layer number of MoS$_2$ of (a) $N$ = 44 and (b) $N$ = 14. The $V_G$ is swept with 2.5 V interval. (c) Current density as a function of $V_G$ at fixed $V_B$ = +0.5 V for $N$ = 44 and 14.

Figure 3. (Color online)
(a, b) Temperature dependence of current density vs $V_G$ at fixed $V_B$ = +0.1 V for graphene/MoS$_2$/Ti vertical FETs with different MoS$_2$ layer number of (a) $N$ = 44 and (b) $N$ = 14. Temperature is swept with 20 K interval. (c, d) The Arrhenius plot of the vertical FETs measured at (c) ON-state ($V_G$ = +50 V) and (d) OFF-state ($V_G$ = -50 V). (e) Change of Schottky barrier height under the modulation of $V_G$.

Figure 4. (Color online)
(a) Amplitude of barrier height modulation $\Delta\varphi$ with respect to the number of MoS$_2$ layers $N$. (b) ON-OFF ratio of the vertical FETs with respect to the number of MoS$_2$ layers $N$. (c) The relation between ON-OFF ratio and $\Delta\varphi$. The dashed line indicate the relationship (ON-OFF ratio) = exp($\Delta\varphi/k_B T$) for $T$ = 300 K.

Figure 1

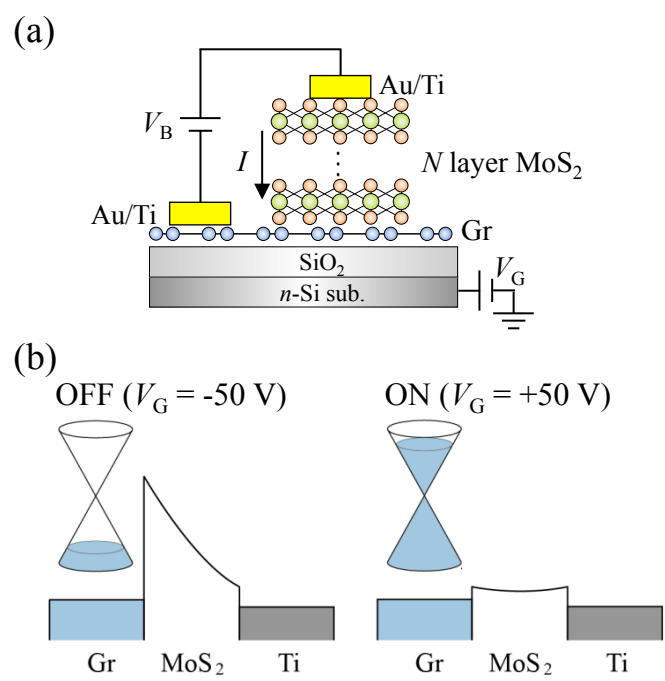

Figure 2

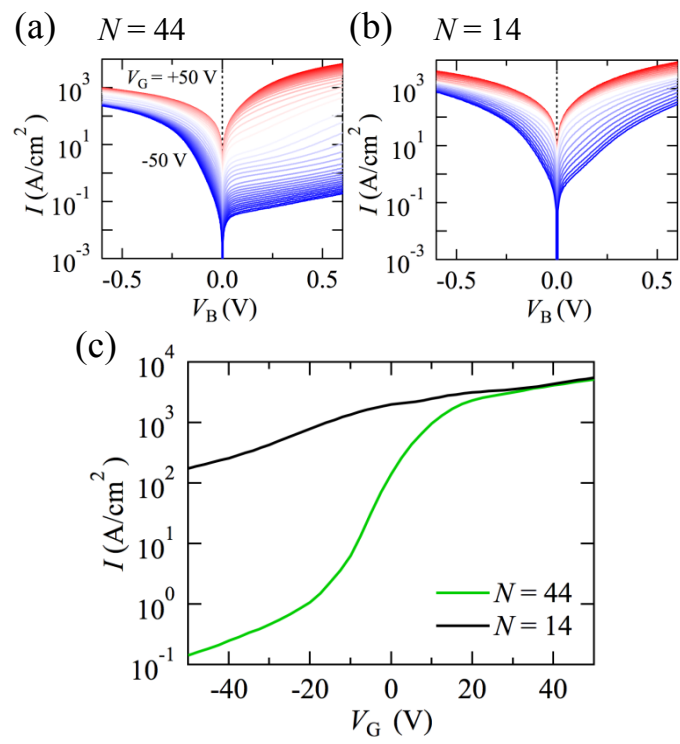

Figure 3

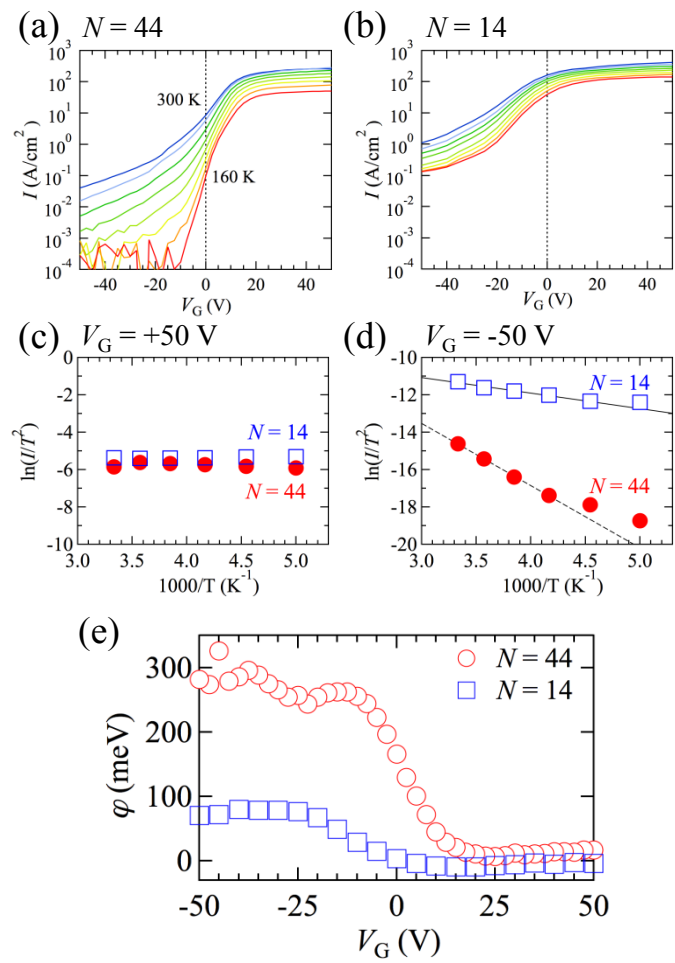

Figure 4

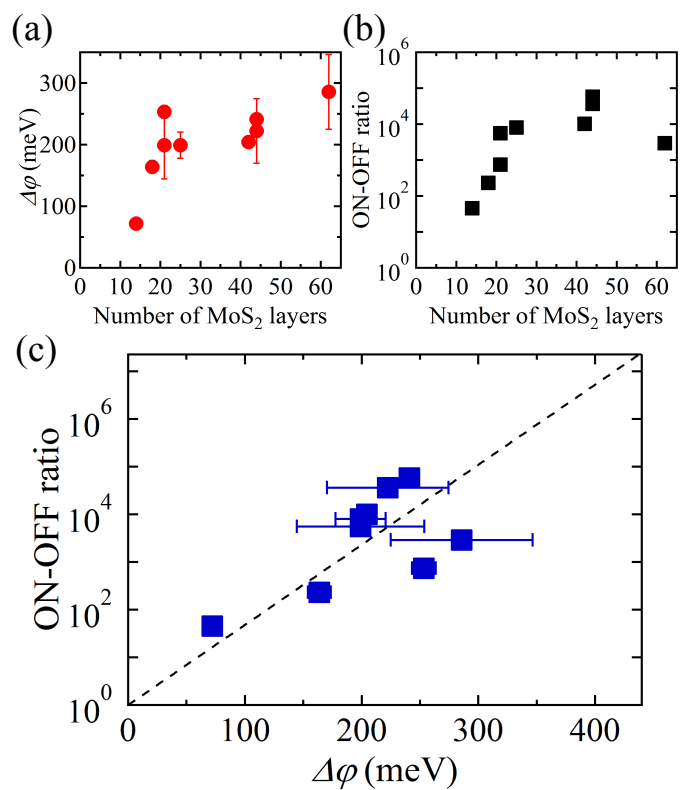